\documentclass[aps,pre,preprint,superscriptaddress]{revtex4}  
\usepackage{graphicx}
\usepackage{amsmath,amssymb,amsfonts}
\usepackage{euscript}

\newcommand{\rhovec}{\boldsymbol{\rho}}

\begin{document}
\title {Avoided intersections of nodal lines}
\author{Alejandro G. Monastra }
\affiliation{Department of Physics of Complex Systems,\\ The Weizmann
Institute of Science, Rehovot 76100, Israel}
\author{Uzy Smilansky}
\affiliation{Department of Physics of Complex Systems,\\ The Weizmann
Institute of Science, Rehovot 76100, Israel}
\author{Sven Gnutzmann}
\affiliation{Institute for Theoretical Physics,\\
Freie Universit\"{a}t Berlin, 14195 Berlin, Germany.}
\begin{abstract}

We consider real eigen-functions of the Schr\"odinger operator in
2-d. The nodal lines of separable systems form a regular grid, and the
number of nodal crossings equals the number of nodal domains. In
contrast, for wave functions of non integrable systems nodal
intersections are rare, and for random waves, the expected number of
intersections in any finite area vanishes. However, nodal lines
display characteristic avoided crossings which we study in the present
work. We define a measure for the avoidance range and compute its
distribution for the random waves ensemble. We show that the avoidance
range distribution of wave functions of chaotic systems follow the
expected random wave distributions, whereas for wave functions of
classically integrable but quantum non-separable wave functions, the
distribution is quite different. Thus, the study of the avoidance
distribution provides more support to the conjecture that nodal
structures of chaotic systems are reproduced by the predictions of the
random waves ensemble.
\end{abstract}
\maketitle
\section{Introduction}
\label{sec:introduction}
The morphology of the nodal sets of wave functions depends crucially
on whether the underlying classical dynamics is integrable or
chaotic. This was first proposed in \cite{miller79} and was followed
by the study of various features of the nodal lines, such as the
distribution of its curvatures \cite{Simmel}. Recently, the counting
statistics of nodal domains for integrable and chaotic systems were
investigated in \cite{BGS}, and it was shown that in the chaotic case,
the statistics follow the predictions derived by assuming that the
wave functions are random superpositions of plane waves
\cite{Berry77,bogoschmidt}. Local effects due to boundary conditions,
and the corresponding modifications of the random wave ensembles were
also studied \cite{Berry02,heller02}.

The interest in the properties of the nodal set is not confined to the
physics literature only. Most of the mathematics literature on this
subject is concerned with solutions of the Helmholtz equation in the
interior of compact domains in $\mathbb{R}^2$ with Dirichlet or
Neumann boundary conditions. Courant \cite{Courant,CH} and later
Pleijel \cite{pleijel} pioneered these studies, and computed an upper
bound on the number of nodal domains. Krahn \cite {krahn24} provided a
lower bound for the area of a nodal domain. Other authors gave
estimates on e.g., the length of the nodal sets \cite{Don-Fef}, and
addressed the general properties of the nodal network \cite{HON}.

\begin{figure}
  \begin{center}
 
    \includegraphics[width=0.85\linewidth]{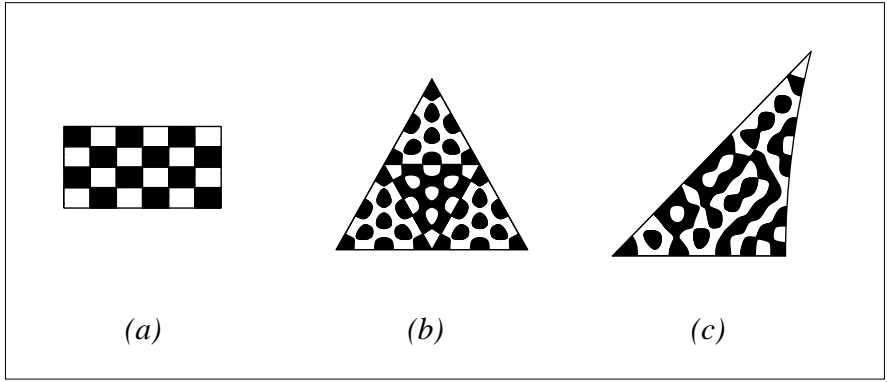}
    \caption{Eigen-functions of 2-d billiards: (a) rectangle (separable
      dynamics); (b) equilateral triangle (integrable, non-separable
      dynamics); (c) Sinai (chaotic dynamics).}
    \label{fig:wavefunctions}
 
  \end{center} 
\end{figure}

Uhlenbeck's theorem \cite{uhlenbeck76} states that the nodal lines of
``generic" wave functions do not intersect (this statement will be
further discussed in the next section). An important class of
exceptions of this rule, are the eigen-functions of separable systems,
where the nodal lines form a grid, and the number of intersections
equals the number of nodal domains. The purpose of the present work is
to provide a {\it quantitative measure} of the degree by which nodal
lines avoid each other. We shall associate the {\it avoidance range}
to each avoided crossing, and will compute their distribution for
the random waves ensemble. The avoidance range vanishes for an
intersection, and therefore, the avoidance range distribution for
separable functions is proportional to a Dirac $\delta$ at zero. We
also show that eigen-functions of classically chaotic systems display
an avoidance range distribution which is similar to the one obtained
for a random waves ensemble, while for an intermediate system, the
avoidance distribution differs substantially from either of the
extreme distributions mentioned above. The nodal structure for a few
representative wave functions can be seen in figure
(\ref{fig:wavefunctions}).

The wave function of a rectangle (a) displays a perpendicular grid of
nodal lines typical of separable systems. An equilateral triangle (b)
is classically integrable, but not separable, and the wave function
shows a few crossings and avoided crossings of the nodal lines. The
domain in (c) is a Sinai billiard where the classical dynamics is
chaotic. With the exception of the boundary crossings, the nodal set
displays a characteristic set of avoided crossings.

The paper is organized in the following way: We shall begin the next
section by discussing the nodal structure of general solutions of the
Helmholtz equation. We shall then identify avoided crossings of the
nodal lines, and define the corresponding avoidance range. The
avoidance range distribution will be written down explicitly. In
section (\ref{sec:avoid-dist}) we shall define the random wave
ensemble and compute explicitly the expected distribution of the
avoidance ranges. The resulting expression will be compared with the
avoidance distributions obtained numerically for high lying
eigen-functions of a chaotic domain. For Dirichlet problems, the
boundary belongs to the nodal set, where nodal intersections occur at
a density which is approximately $2$ intersections per wavelength.
This property cannot be reproduced by the uniform random wave
ensembles, which are expected to account for bulk properties. This
problem, in the context of the avoided crossing distribution, is
discussed at the end of section \ref{sec:avoid-dist}.

\section{Avoided crossings of nodal lines and the avoidance range}
\label{sec:avoid-def}

We consider real solutions of the Helmholtz equation in a domain
$\Omega\in \mathbb{R}^2$ which can be expressed as linear
superpositions of regular solutions of the equation in the entire
plane. They can be expressed in terms of either plane or cylindrical
waves, and the two representations are equivalent. In the plane waves
representation the wave function is written as

\begin{equation}
\Psi({\bf r}) = \sum_{n=-\infty }^{\infty} a_n ~ {\rm e}^{i {\bf k}_n
\cdot {\bf r} }
\label{eq: planewave}
\end{equation}
with wave vectors (${\bf k}_n = -{\bf k}_{-n}\ ,\ \ |{\bf
k}_n|^2=k^2$) directed at an angle $\theta_n$ and coefficients which
ensure that the series (\ref {eq: planewave}) converges absolutely (we
require $a_n^* =a_{-n}$ to render $\Psi({\bf r})$ real). Using the
expansion of a plane wave in cylindrical coordinates we get

\begin{equation}
\Psi({\bf r}) = \sum_{l=-\infty}^{\infty} \alpha_l ~ J_l (k r) ~ e^{i
l \theta} \ ,
\label{eq:cylwaves}
\end{equation}
with

\begin{equation}
\alpha_l =\sum_{n=-\infty}^{\infty} a_n {\rm e}^{il \theta_n}, \ \ \
{\rm and } \ \ \alpha_l^* = (-1)^l \alpha_{-l}\ .
\end{equation}

The infinite sums in both (\ref {eq: planewave}) and (\ref
{eq:cylwaves}) can be truncated, when one considers domains of finite
area. Semiclassical arguments show that the necessary number of terms
is $L \approx {\mathcal L} k / \pi$, where ${\mathcal L}$ is the
 perimeter of the domain. For
reasons which will become clear in the sequel, we prefer to use the
cylindrical wave representation in the present work.

The expansion (\ref {eq:cylwaves}) refers to a particular choice of
the origin. Using Graf's addition theorem \cite {Graf} the
origin can be shifted to ${\bf r}$, and the wave function retains its
form,

\begin{equation}
\Psi ({\bf r} + \rhovec) = \sum_{m} \alpha_m ( {\bf r} ) ~ J_m (k\rho) ~
e^{i m \phi} \ ,
\label {eq:shiftedwf}
\end{equation}
where

\begin{equation}
\alpha_m ({\bf r}) = \sum_{l} \alpha_l ( {\bf 0} ) ~ J_{l-m} (k r) ~
e^{i l \theta} \ ,
\end{equation}
with $\alpha_l( {\bf 0} ) = \alpha_l$, and the angle $\phi$ is
measured from the direction defined by ${\bf r}$. The translated
coefficients $\alpha_m ( {\bf r} ) = \beta_m ( {\bf r} ) + i \gamma_m
( {\bf r}) $ are related to the wave-function and its derivatives
computed at the point ${\bf r}$:

\begin{eqnarray}
\beta_0 ( {\bf r} ) &=& \Psi ( {\bf r} ) \\
 \beta_1 ( {\bf r} ) &=& \frac{1}{k} \Psi_r ( {\bf r} ) \ \ \ \ \ \ \
\ \ \ \ \ \ \ ; \ \ \ \gamma_1 ( {\bf r} ) = -\frac{1}{k r}
\Psi_{\theta} ( {\bf r} ) \nonumber \\
\beta_2 ( {\bf r} ) &=& \Psi ( {\bf r} ) + \frac{2}{k^2} \Psi_{r r} (
{\bf r} ) \ \ ; \ \ \ \gamma_2 ( {\bf r} ) = \frac{2}{k^2 r^2} ~
\left( \Psi_{\theta} ( {\bf r} ) - r \Psi_{r \theta} ( {\bf r} )
\right) \ \nonumber \\
\vspace{5mm} \Psi_{\theta \theta} ( {\bf r} ) &=& -k r ~ \beta_1 (
{\bf r} ) - \frac{k^2 r^2}{2} ~ \left( \beta_2 ( {\bf r} ) + \beta_0 (
{\bf r} ) \right) \nonumber
\label{eq:invderiv}
\end{eqnarray}
In the close vicinity of ${\bf r}$, where $k \rho< 1$, and to second order
in $k\rho$

\begin{eqnarray}
\label{eq:secondord}
\hspace{-15mm}
\Psi ({\bf r} + \rhovec) &\approx& \beta_0({\bf r})\left
(1-\left(\frac{k \rho}{2}\right)^2\right) \\
\hspace{-15mm}
&+&\left | \alpha_1({\bf r})\right |\left ( \frac{k\rho}{2}\right)
\cos(\phi + \phi_1) +\frac{1}{2} \left | \alpha_2({\bf r})\right |
\left (\frac{k\rho}{2}\right)^2 \cos\ 2(\phi + \phi_2) \ , \nonumber
\end{eqnarray}
where $\phi_l$ are the phases of $\alpha_l({\bf r})$. If $\beta_0({\bf
r})=0$, ${\bf r}$ is a nodal point. It cannot be an isolated zero
since the second term vanishes on the line segment through ${\bf r}$
which is oriented at the direction $\frac {\pi}{2}-\phi_1$. Hence, the
nodal set consists of lines. Two nodal lines intersect at ${\bf r}$ if
both $\beta_0({\bf r})=0$ and $\alpha_1({\bf r})=0$, while
$\alpha_2({\bf r}) \ne 0$. The intersection is perpendicular since
$\cos 2(\phi+\phi_2)$ vanishes along two perpendicular lines which
intersect at ${\bf r}$. For the time being we shall continue the
discussion assuming that $\alpha_2({\bf r}) \ne 0$. The more general
case will be commented on at the end of this section.

An {\it avoided crossing} occurs at ${\bf r}$ when $\alpha_1({\bf
r})=0$ and $|\alpha_2({\bf r})| > |\beta_0({\bf r})| > 0$. In other
words, when ${\bf r}$ is a {\it saddle point} of the wave
function. This can be easily seen by writing the equation of the zero
set of (\ref {eq:secondord}) in terms of the local coordinates $\rhovec\
=\ (\xi,\eta)$,

\begin{equation}
1= \xi^2 \left (\frac {\beta_0({\bf r})-|\alpha_2({\bf
r})|}{\beta_0({\bf r})}\right ) + \eta^2 \left( \frac {\beta_0({\bf
r})+|\alpha_2({\bf r})|}{\beta_0({\bf r})} \right )\ .
\label{eq:quadratic}
\end{equation}
This is a hyperbola (ellipse) if $|\alpha_2({\bf r})|$ is larger
(smaller) than $ |\beta_0({\bf r})| $. At an avoided crossing, the
scaled distance between the two branches is

\begin{equation}
z({\bf r}) \equiv k d({\bf r})\ =\ 4 \sqrt{ \frac{ | \beta_0({\bf r})
| }{ | \beta_0({\bf r}) | + |\alpha_2({\bf r})| } } \ .
\label{avoiddist}
\end{equation}
This is the {\it avoidance range} associated with the avoided crossing
at ${\bf r}$.

A few comments are in order:

\noindent {\it i}. At a nodal intersection $z=0$.

\noindent {\it ii}. At a saddle point $|\alpha_2({\bf r})|\ >\
|\beta_0({\bf r})|$, hence $z< 2\sqrt2$.

\noindent {\it iii}. An equivalent expression for $z({\bf r})$ in
terms of $\Psi({\bf r})$ and its Cartesian derivatives reads

\begin{equation}
z({\bf r}) = \left . 4 \sqrt { \frac{ k^2 |\Psi | }{ k^2 | \Psi |
+\sqrt{ 4 \Psi_{xy}^2 + (\Psi_{xx} - \Psi_{yy} )^2 } } }\ \right
|_{\bf r} \ .
\label{avoid1}
\end{equation}

\noindent {\it iv}. Consider an elliptic critical point of the wave
function. In the quadratic approximation the area of the elliptic
nodal domain is
\begin{equation}
{\mathcal A} = 4\pi k^{-2}\ \left. \sqrt{ \frac{ \beta_0^2}{\beta_0^2
-|\alpha_2|^2} } \ \right|_{\bf r} .
\end{equation}
This area is always larger than $4\pi k^{-2}$. Krahn's theorem
\cite{krahn24} gives $j_{0,1}^2 \pi k^{-2}$ as the maximal lower bound
to the area of any nodal domain, where $j_{0,1} \approx 2.405$ is the
first zero of the Bessel function $J_0(x)$. The lower bound $4\pi
k^{-2}$ is smaller but not very far from Krahn's exact value and thus
consistent.

So far we considered the intersections of two nodal lines. However,
higher order intersections may occur. In general, if the first non
vanishing coefficient at ${\bf r}$ is $\alpha_q$, then ${\bf r}$ is a
nodal point of order $q$, where $q$ nodal lines intersect at angles
$\frac {\pi}{q}$. The higher $q$, the more rare are the intersections,
since more conditions are to be satisfied by the coefficients. This
explains Uhlenbeck's theorem \cite {uhlenbeck76} mention above. From
now on we shall discuss the most common intersections with $q=2$, and
comment about the higher order intersections whenever necessary.

Up to now we discussed individual avoided crossings, and defined the
associated avoidance ranges. In the next section we shall consider the
distribution of the avoidance ranges of a wave function in the domain
of its definition, and compute its mean for random waves ensembles.

\section {Avoidance range distributions}
\label {sec:avoid-dist}

The number of the critical points of $\Psi({\bf r})$ in the domain
$\Omega$ is given by

\begin{equation}
\label{eq:ncp}
N_{C} = \int_{\Omega} r {\rm d}r ~ {\rm d}\theta ~ \delta(\Psi_r({\bf
r})) ~ \delta \left( \frac{1}{r}\Psi_{\theta}({\bf r}) \right) |
{\mathcal J} ( {\bf r} ) | \ ,
\end{equation}
with $ {\mathcal J} ( {\bf r} ) = \frac{1}{r} ( \Psi_{rr}({\bf r})
\Psi_{\theta \theta}({\bf r}) - \Psi_{r\theta}^2 ({\bf r}) ) $ the
Jacobian. Using (\ref{eq:invderiv}) we get

\begin{equation}
\label{eq:ncp2}
N_{C} = \frac{k^2}{4} \int_{\Omega} r {\rm d}r ~ {\rm d}\theta ~
\delta(\beta_1({\bf r})) ~ \delta(\gamma_1({\bf r})) ~ \left|
|\alpha_2({\bf r})|^2 -\beta_0^2({\bf r}) \right| \ .
\end{equation}
To count the saddle points, we add the restriction $|\alpha_2({\bf
r})|^2 -\beta_0^2({\bf r}) >0$ and obtain

\begin{equation}
\hspace{-15mm}
N_{S} = \frac{k^2}{4} \int_{\Omega} r {\rm d}r ~ {\rm d}\theta ~
\delta(\beta_1({\bf r})) ~ \delta(\gamma_1({\bf r})) ~ (
|\alpha_2({\bf r})|^2 -\beta_0^2({\bf r}) ) ~ \Theta ( |\alpha_2({\bf
r})|^2 -\beta_0^2({\bf r}) ) \ .
\label{eq:nsp}
\end{equation}
Combining (\ref {eq:nsp}) and (\ref {avoiddist}), the number of
avoided crossings with avoidance ranges less than $z$ is

\begin{equation}
\tilde{\mathcal I} (z) = \frac{k^2}{4} \int_{\Omega} r {\rm d}r {\rm
d}\theta ~ \delta(\beta_1({\bf r})) \delta(\gamma_1({\bf r})) (
|\alpha_2({\bf r})|^2 -\beta_0^2({\bf r}) ) \Theta ( |\alpha_2({\bf
r})|^2 -\beta_0^2({\bf r}) )\Theta ( z - z({\bf r})) \ ,
\label{eq:Iofz}
\end{equation}
and the fraction of the total number is $ {\mathcal I}(z) =
\tilde{\mathcal I} (z) / N_S $. This counting function, and its
associated density ${\mathcal P}(z) \equiv \frac {{\rm d} {\mathcal
I}(z) }{{\rm d} z}$ are the distributions which characterize the
avoided crossings of the nodal set. The range of $z$ is
$[0,2\sqrt2]$. ${\mathcal I}(z)$ is normalized such that it takes the
value $1$ at $z=2\sqrt 2$. One can easily check that an avoided
crossing of order $q$ is counted in (\ref {eq:Iofz}) with a
multiplicity $q-1$. The number (including multiplicity) of nodal
crossings provides the value of ${\mathcal I}(0)$. The effective
multiplicity of nodal crossings of the boundary is reduced by a factor
one half.

\begin{figure}
  \begin{center}
    \includegraphics[width=0.85\linewidth]{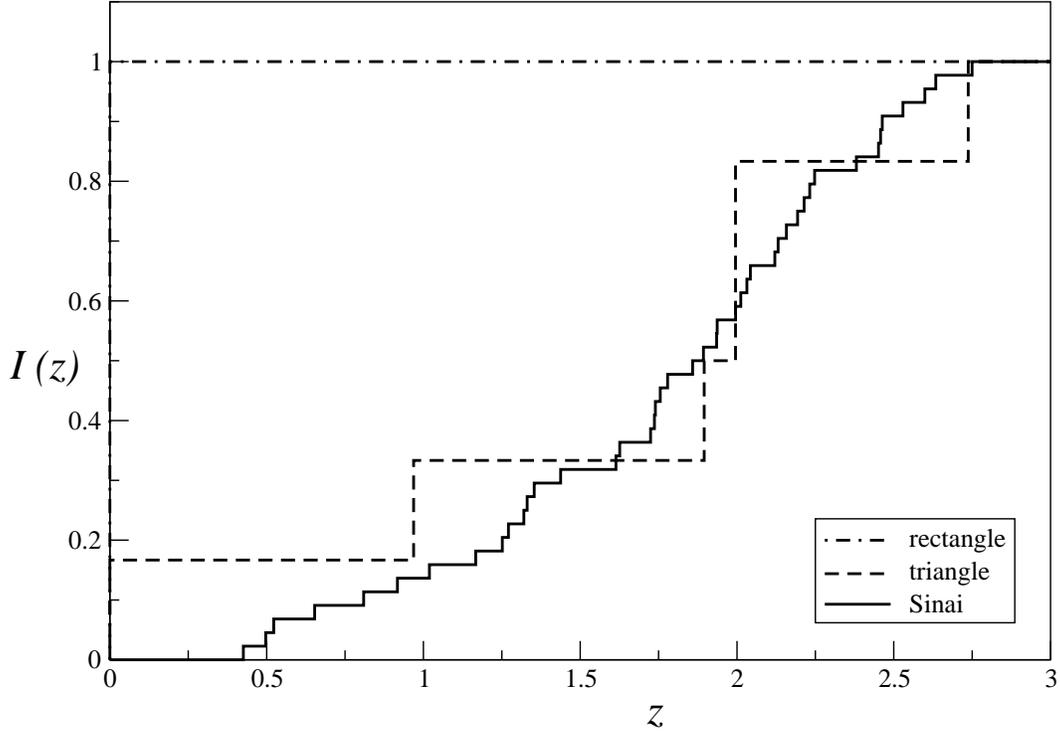}
    \caption{Normalized cumulative histograms of the avoidance ranges
      ${\mathcal I} (z)$ for the three wave functions plotted in figure 1:
      rectangle (dash-point line); triangle (dashed line); Sinai (full
      line). }
    \label{fig:Iofz}
  \end{center}
\end{figure}

Numerically computed ${\mathcal I} (z) $ are shown in figure \ref
{fig:Iofz} for the three wave functions plotted in figure \ref
{fig:wavefunctions}. The computed ${\mathcal I} (z) $ do not take into
account the nodal crossings of the boundary. For the separable
billiard all the saddle points have zero avoidance range, and the
function ${\mathcal I} (z) $ is trivially equal to one in all the $z$
range. In the triangular billiard there are still some saddle points
with zero avoidance range (nodal crossings), but most of them have a
finite avoidance range. For the chaotic billiard, all the calculated
saddle points have a finite avoidance range.

\subsection{Avoidance range distribution for the random waves ensemble}
\label {subsec:randwaves}

One of the main goals of the present work is to show that the
properties of the nodal set of chaotic billiards, as detected by the
distribution of avoidance ranges, are reproduced by the distributions
computed for the random wave ensembles. Because wave functions are
subject to boundary conditions, it is expected that the predictions of
the isotropic random wave ensemble used e.g. in \cite
{BGS,bogoschmidt}, are relevant only to the bulk of the domain, and
will do poorly in the $\lambda=\ (\frac {2\pi}{k})$ vicinity of the
boundary. This was observed and discussed in \cite {BGS}.

The isotropic random wave ensemble is the ensemble of wave functions
(\ref {eq:cylwaves}) where the real parameters $\beta_l$ and
$\gamma_l$ are independent, identically distributed random Gaussian
variables with zero mean and unit variance for all $|l|\ge 1$. Because
$\gamma_0=0$ the variance of $\beta_0$ is twice that of all the
others. The ensemble average of a function $f$ will be denoted by
$\langle f \rangle$. The local coefficients $\alpha_l({\bf r})
=\beta_l({\bf r}) +i\gamma_l({\bf r}) $ were derived from the original
ones by a unitary transformation. Hence, they are also independent
identically distributed Gaussian variables, and the ensemble averages
of the number of critical points (\ref {eq:ncp}), the number of saddle
points (\ref {eq:nsp}) and the mean distribution of avoidance ranges
(\ref {eq:Iofz}) can be computed by considering Gaussian integrations
with respect to the variables $ \beta_0 , \beta_1 ,\gamma_1, \beta_2$
and $\gamma_2$. A straightforward integration gives,

\begin{equation}
\langle N_{C} \rangle = \frac{k^2 |\Omega|}{2 \pi \sqrt{3}} \ \ \ ;\ \ \
\langle N_{S} \rangle = \frac{k^2 |\Omega|}{4 \pi \sqrt{3}} \ ,
\label{meansp}
\end{equation}
where $|\Omega|$ is the area of the domain. The average number of
saddle points is a half of the number of critical points. The other
half are the points were the wave function has either a minimum or a
maximum.

The mean number of saddle points with an avoidance range less than or
equal to $z$ can be also computed,

\begin{equation}
\langle \tilde{\mathcal{I}} (z) \rangle = \frac{k^2 |\Omega| }{4 \pi}
\frac{3 ~ z^2 ~ (16 - z^2)^2 }{ (512 - 64 z^2 + 3 z^4 )^{3 /2} } \ ,
~~~~ 0 < z < 2 \sqrt{2} \ .
\label{intprobz}
\end{equation}
The normalization gives trivially the mean number of saddle points in
the ensemble. Therefore we can define the probability

\begin{equation}
{\mathcal{I}}_{r.w.} (z) = \frac{ \langle \tilde{\mathcal{I}} (z) \rangle }{
\langle N_{S} \rangle } = \frac{ 3 \sqrt{3} ~ z^2 ~ (16 - z^2)^2 }{
(512 - 64 z^2 + 3 z^4 )^{3 /2} } \ , ~~~~ 0 < z < 2 \sqrt{2} \ ,
\label{izrw}
\end{equation}
with a corresponding density

\begin{equation}
{\mathcal{P}}_{r.w.} (z) = \frac{ 6144 \sqrt{3} ~ z ~ (8 - z^2) ~ (16 - z^2)
}{ (512 - 64 z^2 + 3 z^4 )^{5/2} } \ , ~~~~~~ 0 < z < 2 \sqrt{2} \ .
\end{equation}

\begin{figure}
  \begin{center}
    \includegraphics[width=0.8\linewidth]{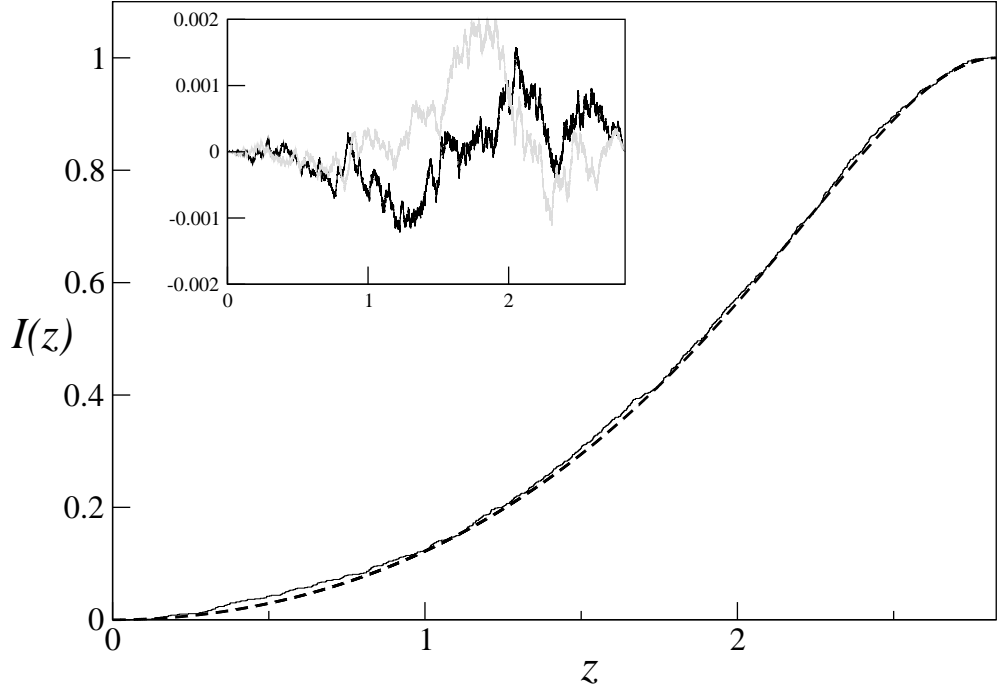}
    \caption{Counting function ${\mathcal I} (z)$ for the $n=2000$
      eigen-state of the Sinai type billiard (full thin line), compared with
      the random wave prediction (\ref{izrw}) (dash line). Inset: difference
      between the average counting function over the $n$ = (1800, 2000)
      eigen-states and the random wave prediction (dark line); the same for
      $n$ = (2200, 2400) eigen-states (light line). }
    \label{fig:Irw}
  \end{center}
\end{figure}

These results served to test the conjecture that the bulk properties
of the nodal sets of chaotic wave functions are reproduced by the
predictions of the isotropic random wave ensemble. For this purpose we
computed numerically the first 2400 eigen-functions of the Sinai
billiard shown in figure \ref{fig:wavefunctions}.c. For each
wave-function the critical points were found by a numerical
search. The saddles which correspond to boundary intersections were
excluded. The avoidance range (\ref{avoid1}) was computed for each
saddle. In figure \ref{fig:Irw} we plot the cumulative histogram of
the avoidance ranges for the 2000th eigen-state in full line (1102
saddle points), compared with formula (\ref{intprobz}) in dashed
line. Even for a single eigen-state the agreement is very
good. Averaging the avoidance range distributions over a group of
neighboring eigen-states the numerical histogram and the theoretical
curve approximately coincide. The inset in figure \ref{fig:Irw}
shows the differences between the random wave prediction and the mean
avoidance range distributions computed for two groups of
eigen-states. The difference is small, and shows no systematic
deviations.

\begin{figure}
  \begin{center}
    \includegraphics[width=0.8\linewidth]{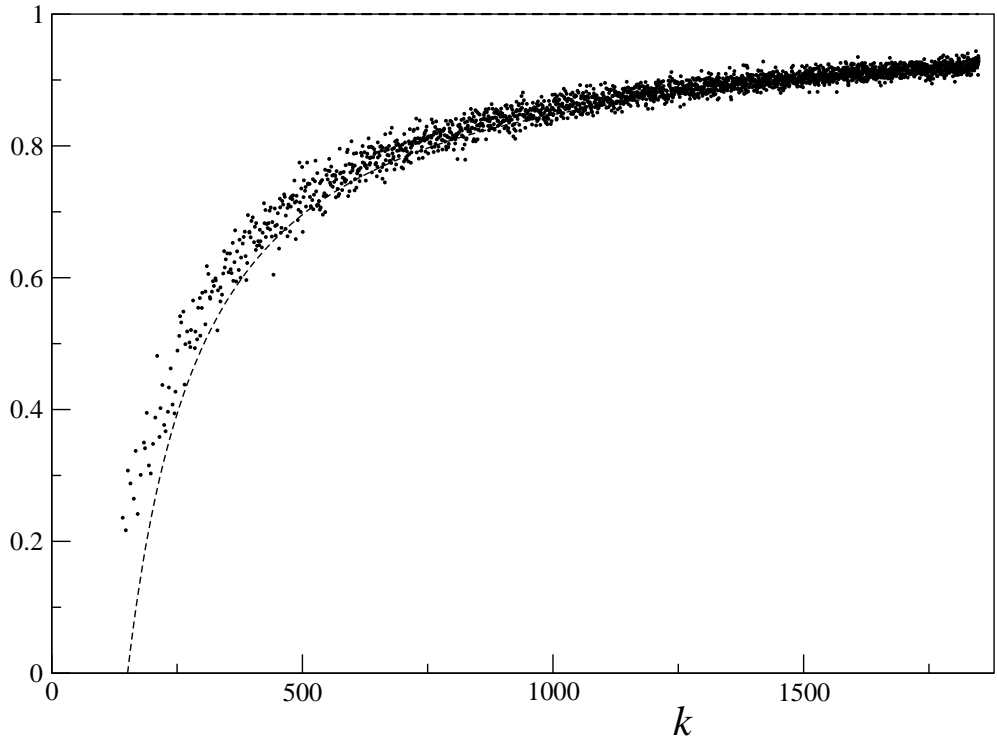}
    \caption{Number of saddle points for the first 2400 eigen-states of
      the Sinai billiard, normalized by the isotropic random wave prediction
      $k^2/(4 \pi \sqrt{3})$, as a function of the wave number $k$. Dashed
      line: equation (\ref{eq:nspdirich}) for the anisotropic random wave
      ensemble (\ref{eq:pwaves}).}
    \label{fig:nsp}
  \end{center}
\end{figure}

Another distribution which we compared to the prediction of the random
waves ensemble is the number of saddle points. Normalizing the number
of saddle points by the prediction of the isotropic random wave
ensemble, we observe as $k$ increases that the numerical computation
approach the predicted value (see figure \ref {fig:nsp}).

The systematic deviation observed at finite $k$ is due to boundary
effects. Following \cite{Berry02} we compute the effect of an infinite
straight Dirichlet line, which is reproduced by the wave ensemble

\begin{equation}
\Psi ( {\bf r} ) = 2 ~ \sum_{n = 1}^{\infty} c_n ~ \sin (n \theta) ~
J_{n} (k r) \ ,
\label{eq:pwaves}
\end{equation}
where the $c_n$ are real coefficients taken as independent random
Gaussian distributed variables.

Not entering into the details of the calculation, the computed mean
density of saddle points approach the bulk expression (\ref {meansp})
as the distance from the Dirichlet line increases. The integrated
density in the perpendicular direction shows a global deficiency of
saddle points relative to the bulk value. It diverges logarithmically
as a function of the distance from the Dirichlet line \cite
{Berry02}.

We compared the results of the ensemble (\ref{eq:pwaves}) to the
number of saddle points counted for highly exited eigen-states of a
chaotic billiard with sufficiently smooth boundaries. The integral
along the boundary multiplies the density by the perimeter of the
billiard ${\mathcal{L}}$. The perpendicular integral must be truncated in
view of the logarithmic divergence mentioned above. A sensible choice
of the truncation distance is $R = \sqrt{|\Omega|}/2$. The resulting
estimate for the mean number of saddle points is

\begin{equation}
\langle N_{S} \rangle \approx \frac{ k^2 |\Omega| - k {\mathcal{L}} (
\sigma_1 \log (k R) + \sigma_2 ) }{4 \pi \sqrt{3} } \
. \label{eq:nspdirich}
\end{equation}
with $\sigma_1 \approx 0.014$ and $\sigma_2 \approx 2.0$. The
deficiency of saddle points is explained by the effect of the
Dirichlet boundaries that affects the statistics. The dashed line in
figure \ref{fig:nsp} represents the equation (\ref{eq:nspdirich}) and
the agreement is definitely improving. However, the domain of low
values $k \ < \ 500$ is not in complete agreement. This deficiency can
be associated with the corners of the billiard, to its curvature, or
to the finiteness of the Dirichlet line that the ensemble
(\ref{eq:pwaves}) can not reproduce. These possible causes will be
studied elsewhere.

In conclusion we can say that the properties of the nodal set of
chaotic wave functions which were investigated in this work, are very
well reproduced by the isotropic random waves ensemble in the
semiclassical domain. These findings are consistent with previous
works on the subject, and add support to the random waves conjecture.

\section {Acknowledgments}
This work was supported by the Minerva Center for non-linear Physics
at the Weizmann Institute and by an ISF research grant. AGM
acknowledges a post doctoral fellowship from the European Network on
{\it Mathematical aspects of Quantum Chaos} which support his stay at
the Weizmann Institute.


\end{document}